\documentclass[prb,twocolumn,superscriptaddress,showpacs]{revtex4}
\usepackage{mathrsfs}
\usepackage{amsmath,amsfonts,amssymb}
\usepackage{epsfig}
\usepackage{graphicx}
\usepackage{wasysym}
\usepackage{bbm}
\usepackage{psfrag}
\usepackage{color}
\usepackage{pstool}
\usepackage{braket}
\usepackage[dvips, bookmarks, colorlinks=true, plainpages = false, citecolor = blue, linkcolor = blue, urlcolor = blue, filecolor = blue]{hyperref}
\newcommand \be{\begin{equation}}
\newcommand \ee{\end{equation}}
\newcommand \bes{\begin{equation*}} 
\newcommand \ees{\end{equation*}}
\newcommand \bea{\begin{eqnarray}}
\newcommand \eea{\end{eqnarray}}
\newcommand \bsea{\begin{subequations}\begin{eqnarray}} 
\newcommand \esea{\end{eqnarray}\end{subequations}}
\newcommand \beas{\begin{eqnarray*}} 
\newcommand \eeas{\end{eqnarray*}}
\newcommand \bfg{\begin{figure}}
\newcommand \efg{\end{figure}}
\newcommand \bfgs{\begin{figure*}} 
\newcommand \efgs{\end{figure*}}
\newcommand \bwt{\begin{widetext}}
\newcommand \ewt{\end{widetext}}

\def\pmat#1{\left(\begin{matrix}#1\end{matrix}\right)}

\begin{document}
\title{First and Second Order Topological Phases on Ferromagnetic Breathing Kagome Lattice }
\author{Arghya Sil}\email{arghyasil36@gmail.com}
\affiliation{Department of Physics, Jadavpur University, 188 Raja Subodh Chandra Mallik Road, Kolkata 700032, India}
\author{Asim Kumar Ghosh}\email{asimkumar96@yahoo.com}
\affiliation{Department of Physics, Jadavpur University, 188 Raja Subodh Chandra Mallik Road, Kolkata 700032, India}

\begin{abstract}
In this work, topological properties of a ferromagnetic 
Heisenberg model on a breathing kagome lattice are
investigated extensively in the presence of 
 Dzyaloshinskii-Moriya interaction. 
While the kagome ferromagnet hosts only a single first order 
topological phase, 
the breathing kagome system exhibits multiple first and second order 
topological phases along with their coexistence. 
Magnon dispersion relation is obtained by using linear spin wave theory. 
Flat band and Dirac cones are obtained in the absence of 
 Dzyaloshinskii-Moriya interaction. 
A topological 
phase diagram is presented where several first and second order phases 
as well as their overlap are identified. 
Values of thermal Hall conductivity for all the first order 
phases are obtained. Distinct first order phases 
are characterized by different sets of Chern numbers in association with the 
necessary chiral edge states in accordance to the first order 
bulk-boundary-correspondence rule. Second order phase is characterized by 
polarization along with the emergence of corner states. 
Violation of the 
second order bulk-corner-correspondence rule has been noted in some 
regions. 
\end{abstract}
\maketitle
\date{today}

\section{Introduction}

Topological states of matter have become one of the most studied topics
for several years. A topological insulator (TI) is characterized by 
gapped bulk states and gapless boundary states or edge states. Time reversal
symmetry (TRS) breaking TI's are known as Chern insulator\cite{Haldane}, where each energy
band is associated with a definite Chern number\cite{TKNN}, while time reversal invariant
TI's are characterized by a nontrivial $Z_2$ invariant \cite{Kane}. Bulk-boundary-correspondence
(BBC) rule \cite{Hatsugai1,Hatsugai2} determines the relation between the bulk and boundary properties
of such systems and gives topological protection to the edge states. Also, there
are topological crystalline insulators \cite{Furusaki}, where mirror Chern number 
acts as the topological index. 

Recently, the concept
of higher order TI's (HOTI) \cite{BBH1,BBH2,Peng,Langbehn,Song,Schindler} has been 
introduced where a $d$ dimensional 
$n$-th order TI shows $(d\!-\!n)$ dimensional boundary states contrary to the conventional
or first order TI's when $n=1$. For instance, a two-dimensional second order
TI (SOTI) will show zero-dimensional corner states but will not show one-dimensional
gapless edge states \cite{Ezawa,Sen,Hatsugai,Geier,Nori,Trebst,Wang,Chen,Fukui}.
In those HOTI's, the conventional BBC rule may not 
be applicable. Obviously, different types of topological invariants like polarization \cite{Ezawa},
$Z_Q$ Berry phase \cite{Hatsugai,Hatsugai3}, nested Wilson loop \cite{BBH2},
mirror Chern number \cite{Sen} etc have been introduced 
depending on the symmetry of the system
to characterize the topological property of the bulk. 
Origin of lower dimensional edge states
can be attributed to quantization of dipole or quadrupole moment 
as observed in two-dimensional phononic and electric quadrupole 
topological insulators \cite{BBH1,Serra,Imhof}. 

In recent times, besides electronic TI, topological magnon insulators (TMI) \cite{Li} are being
studied, where the quasiparticle excitation (magnon) is bosonic in nature. It has been 
known that the topological nature of a system is independent of the statistics of the 
quasiparticles. Topological magnons are found before in 
honeycomb lattice \cite{Owerre1,Owerre2}, kagome lattice \cite{Sen2,Li}, Lieb lattice 
\cite{Cao}, etc. Those topological phases have been experimentally 
observed in kagome ferromagnets, Lu$_2$V$_2$O$_7$ \cite{Tokura,Lee}, 
Cu[1,3-benzenedicarboxylate (bdc)] \cite{Chisnell}, and honeycomb ferromagnet, CrI$_3$ \cite{Chen2}.
Topological magnons give rise to thermal magnon Hall effect (MHE),
in which a temperature gradient transports a dissipationless heat 
current which has been verified experimentally
\cite{Ong}. Generally, in those spin systems, ferromagnetic (FM) Heisenberg model is 
considered where Dzyaloshinskii-Moriya interaction (DMI) is incorporated to 
trigger non-zero berry curvature. Spin models without DMI 
have also been shown to possess non-trivial topology. For example, 
FM Heisenberg models with Kitaev and spin-anisotropic interactions (HKSA)
are found to host a number of topological phases \cite{Joshi,Deb}. 
Thus, topological magnons have promising applications in the field of
dissipationless spin transport, magnon spintronics and magnetic data storage. 

On the other hand, HOTI's have been studied so far in fermionic systems 
in terms of tight-binding models on square and cubic lattice \cite{BBH1,BBH2,Peng,Schindler},
breathing kagome \cite{Ezawa,Zhang}, photonic systems \cite{Chen,Dong}, non-Hermitian systems
\cite{Nori}, etc. In addition, Kitaev model on Shastry-Sutherland lattice and 
magnetic vortex model on kagome lattice exhibit HOTI phases \cite{Trebst,Wang}. 
Higher order topological Mott insulating phase has been 
demonstrated in a Hubbard model on the kagome lattice, 
where the topological state is characterized by $Z_3$ 
spin-Berry phase \cite{Kudo}. 
Besides, SOTIs have been experimentally realized using quantized
dipole or quadrupole polarization \cite{Serra,Imhof} and implemented in
mechanical systems \cite{Serra}, electrical circuits \cite{Ezawa2}, microwave systems 
\cite{Peterson}, photonic \cite{Hassan} and phononic crystals \cite{Zhang}.
However, no report on higher order topological phase in the FM Heisenberg systems 
is available till date. 

In this work, we focus on the realization of second order topological 
magnon insulating (SOTMI) phase in a spin 
system with and without DMI. Here, FM Heisenberg
model is formulated on the breathing kagome lattice 
in the presence of DMI along the nearest neighbor (NN) bonds. 
The system reveals the existence of simultaneous first and second order 
TMI phases in different parameter regimes when DMI is non-zero. 
When DMI is zero, only second order TMI can be realized 
as the Chern number ($C$), the first order topological invariant, 
is always zero for all the bands. 
In other case, polarization is used as the bulk topological 
index to characterize the HOTI phase due to
the mirror symmetry of the system \cite{Ezawa}. 
So, polarization plays the crucial role 
to distinguish between the nontrivial 
and trivial SOTI phase in the same way $C$ distinguishes 
between the nontrivial and
trivial TMI phases in case of first order. 
One dimensional gapless edge states are found for 
nontrivial TI phase, while gapped edge states along with
zero dimensional corner states are found for nontrivial SOTMI phase.
For certain values of DMI strength, both type of phases are found to exist
simultaneously. Transition between different topological phases are shown 
in the parameter-space. In addition, thermal Hall conductivity is
calculated for all the TMI phases.

The article is organized in the following way.
In section \ref{model}, breathing kagome lattice is 
described and the linear spin-wave Hamiltonian is formulated. 
We describe the topological
phases for zero DMI strength in the following 
section \ref{properties}. Topological phases
for non-zero DMI strength are explained in the 
subsequent section \ref{DM}.
The values of thermal Hall conductivities are available 
in  section \ref{THCD}.
Finally, section \ref{summary} contains the discussion 
along with the summary of the results. 
\section{Formulation of Heisenberg Hamiltonian with DM Interaction}
\label{model}
\begin{figure}[t]
\includegraphics[width=8cm,height=5cm]{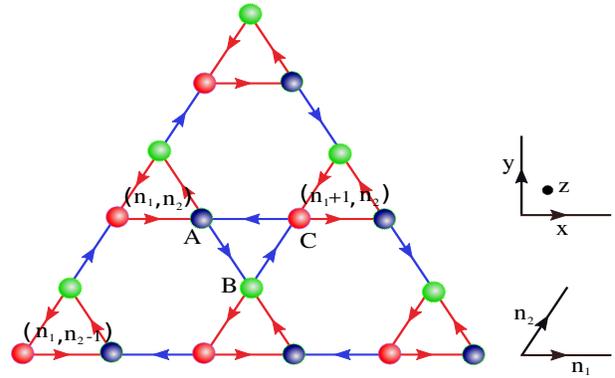}
\caption{(color online) A triangular replica of breathing kagome lattice is shown. Three sites A, B
  and C are denoted by green, blue and red spheres, respectively. The notation of the unit cells is
  also shown. The Heisenberg interaction strength is $J_1$ for the red lines (upward triangles)
  and $J_2$ for the blue lines (downward triangles). Considering DMI along $z$ direction,
  coupling strength of two sites along the arrow will be $J_1 \,(J_2) + i D_1 \,(D_2)$ and opposite to the arrow
  will be $J_1 \,(J_2) - i D_1 \,(D_2)$. The lattice vectors,  $\mathbf{n_1}=(1,0),\mathbf{n_2}=(1/2,\sqrt{3}/2)$ are shown
in the side diagram.}
\label{lattice}
 \end{figure}

A FM Heisenberg Hamiltonian is formulated 
on the breathing kagome lattice with 
DMI along NN bonds. Breathing kagome lattice 
is composed of three identical triangular sublattices. As a result,  
the unit cell comprises of three sites A, B and C
forming a downward triangle (Fig \ref{lattice}). The spin operators on those three sites
are denoted by $\mathbf{S^a_n}, \mathbf{S^b_n} $ and 
$\mathbf{S^c_n}$, respectively. The coordinates of a unit cell are
denoted by $\textbf{n}=(n_1,n_2)$. So, the Hamiltonian of this system can be written as 
\bea
H=H_{\rm NN}+H_{\rm mag}+H_{\rm DM},
\eea
where
\bea
H_{\rm NN}&\!\!=\!\!&-J_{\alpha}\sum_{\langle n,n'\rangle}\!\!\left(\mathbf{S^a_n}\cdot\mathbf{S^b_{n'}}
+\mathbf{S^b_n}\cdot \mathbf{S^c_{n'}}+\mathbf{S^c_n}\cdot \mathbf{S^a_{n'}}\right),\nonumber\\
H_{\rm mag}&=&-h\sum_n\left(S^{az}_n+S^{bz}_n+S^{cz}_n\right),\nonumber\\
H_{\rm DM}\!\!&=&\!\!-D_{\alpha} \!\!\sum_{\langle n,n'\rangle}\!\!\!\left(\mathbf{S^a_n}\times\mathbf{S^b_{n'}}\!+\!
\mathbf{S^b_n}\times \mathbf{S^c_{n'}}\!+\!\mathbf{S^c_n}\times \mathbf{S^a_{n'}}\right)\cdot\hat{z}.\nonumber
\eea
$J_{\alpha},\,\alpha=1\,(2)$ is the NN exchange interaction strength between upward (downward) 
triangles and $D_{\alpha},\,\alpha=1\,(2)$ is the DMI strength, pointing towards $z$ ($-z$) direction, between 
upward (downward) triangles. $\langle \cdot\rangle$ denotes the summations over NN pairs.
Considering the FM case, we fix $J_{\alpha} > 0$ throughout the paper. 
$h=g\mu_B{\mathcal H}$, where ${\mathcal H}$
is the strength of the external magnetic field along $z$ direction, which helps to align the
localized spins ferromagnetically along $z$ direction when $h$ is 
assumed greater than zero.

Now, based on the classical ground state in which all the spins point along the $+z$ direction,
we obtain FM magnon dispersion relation by expressing the spin operators,  $\mathbf{S^\eta_n}$, 
in terms of bosonic
creation ($\eta^{\dagger}$) and annihilation operators ($\eta$) by using the standard Holstein-Primakoff (HP)
transformation:
\begin{equation}
  \begin{aligned}
    S^{\eta z}_{\mathbf{n}}=S-\eta^{\dagger}_{\mathbf{n}}\eta_{\mathbf{n}},\,
    S^{\eta +}_{\mathbf{n}}\simeq\sqrt{2S}\,\eta_{\mathbf{n}},\,
    S^{\eta -}_{\mathbf{n}}\simeq\sqrt{2S}\,\eta^{\dagger}_{\mathbf{n}},\nonumber
  \end{aligned}
\end{equation}
where $\eta=a,b$ and $c$ for the respective sublattices A, B and C.  
 $S^{\eta\pm}_{\mathbf{n}}=
S^{\eta x}_{\mathbf{n}} \pm iS^{\eta y}_{\mathbf{n}}$.
Now, following linear spin wave theory (LSWT) and using Fourier transformation of the operators
in the form 
$\eta_{\mathbf{n}}=\frac{1}{\sqrt N}\sum_{\mathbf{k}}\eta_{\mathbf{k}}e^{i{\mathbf{k}}\cdot{\mathbf{n}}}$  
($N$ is the total number of unit cells in the lattice),   
the Hamiltonian in the momentum space can be written as
\bea
H=E_0+H_{SW},
\eea
where $E_0=-h\sum_nS-3(J_1+J_2)\sum_{\langle n,n'\rangle}S^2$, is the 
classical ground state energy. 
$H_{SW}$ can be written as (retaining terms only up to second order in bosonic operators)
\be H_{SW}=S\sum_k {\psi_k}^{\dagger} {\mathcal M}(\textbf{k}) \psi_k, \ee
 where $\textbf{k} =(k_x,k_y)$, $\psi_{\textbf{k}} = \left(c_{\textbf{k}},
 a_{\textbf{k}},b_{\textbf{k}}\right) $  and ${\mathcal M}(\textbf{k})$ 
is a $3\times 3$ matrix which is given by
 \begin{equation}
   \begin{aligned}
     {\mathcal M}(\textbf{k})=\pmat{m_{11}&m_{12}&m_{13}\\m^\ast_{12}&m_{22}&m_{23}
\\m^\ast_{13}&m^\ast_{23}&m_{33}}, 
   \end{aligned}
   \end{equation}
 with the components, $m_{ij}$, given by
  \begin{equation}
    \begin{aligned}
      m_{11}&=m_{22}=m_{33}=2(J_1+J_2)+h/S,\\
      m_{12}&=-(J_1+iD_1)-(J_2+iD_2)e^{-ik_1},\\
      m_{13}&=-(J_1-iD_1)-(J_2-iD_2)e^{-ik_2},\\
      m_{23}&=-(J_1+iD_1)-(J_2+iD_2)e^{i(k_1-k_2)},
 \end{aligned}
   \end{equation}     
  where $k_1=\mathbf{k}\cdot\mathbf{n_1}=k_x$ and $k_2=\mathbf{k}\cdot\mathbf{n_2}=k_x/2+\sqrt{3}k_y/2$.
Magnetic field only appears in each of the diagonal terms of 
${\mathcal M}(\textbf{k})$ with a fixed value, $h/S$, which 
means that topological properties of this system are totally 
insensitive to the value of ${\mathcal H}$. 
We have assumed a very small positive value of $h$ 
only to ensure the FM ground state.   
  As the Hamiltonian $H_{SW}$ is number conserving, the magnon dispersion relation can be obtained
  by diagonalizing it. The results are valid for any value of $S$, 
while accuracy increases with the magnitude of $S$. 
  
  \begin{figure*}
   \centering
\includegraphics[width=5cm,height=6cm,trim={3.0cm 1.0cm 3.0cm 1.5cm}]{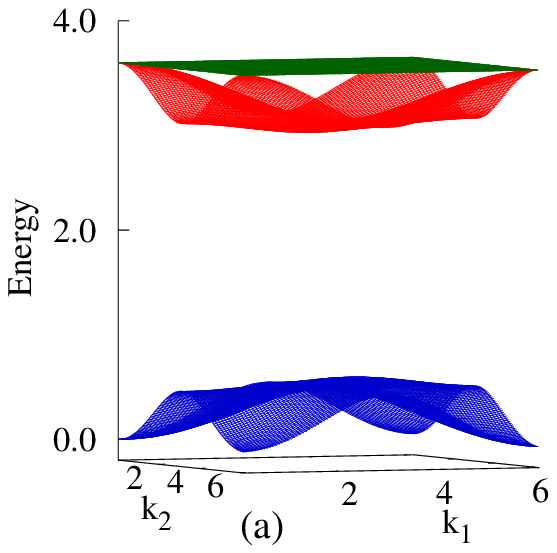}
\includegraphics[width=5cm,height=6cm,trim={3.0cm 1.0cm 3.0cm 1.5cm}]{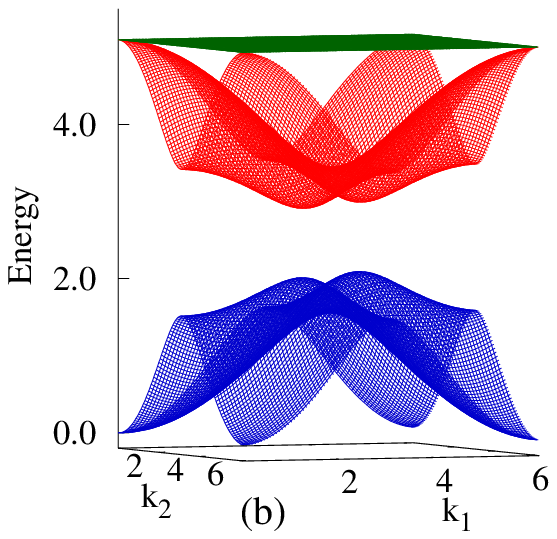}
\includegraphics[width=5cm,height=6cm,trim={3.0cm 1.0cm 3.0cm 1.5cm}]{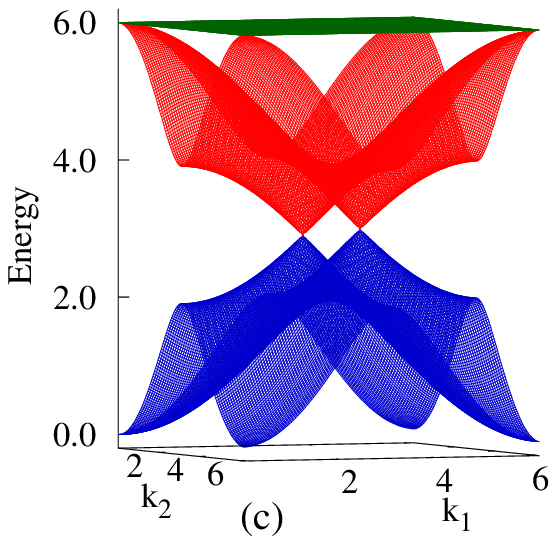}
\includegraphics[width=5cm,height=4cm,trim={0.0cm 0.0cm 0.0cm 0.5cm}]{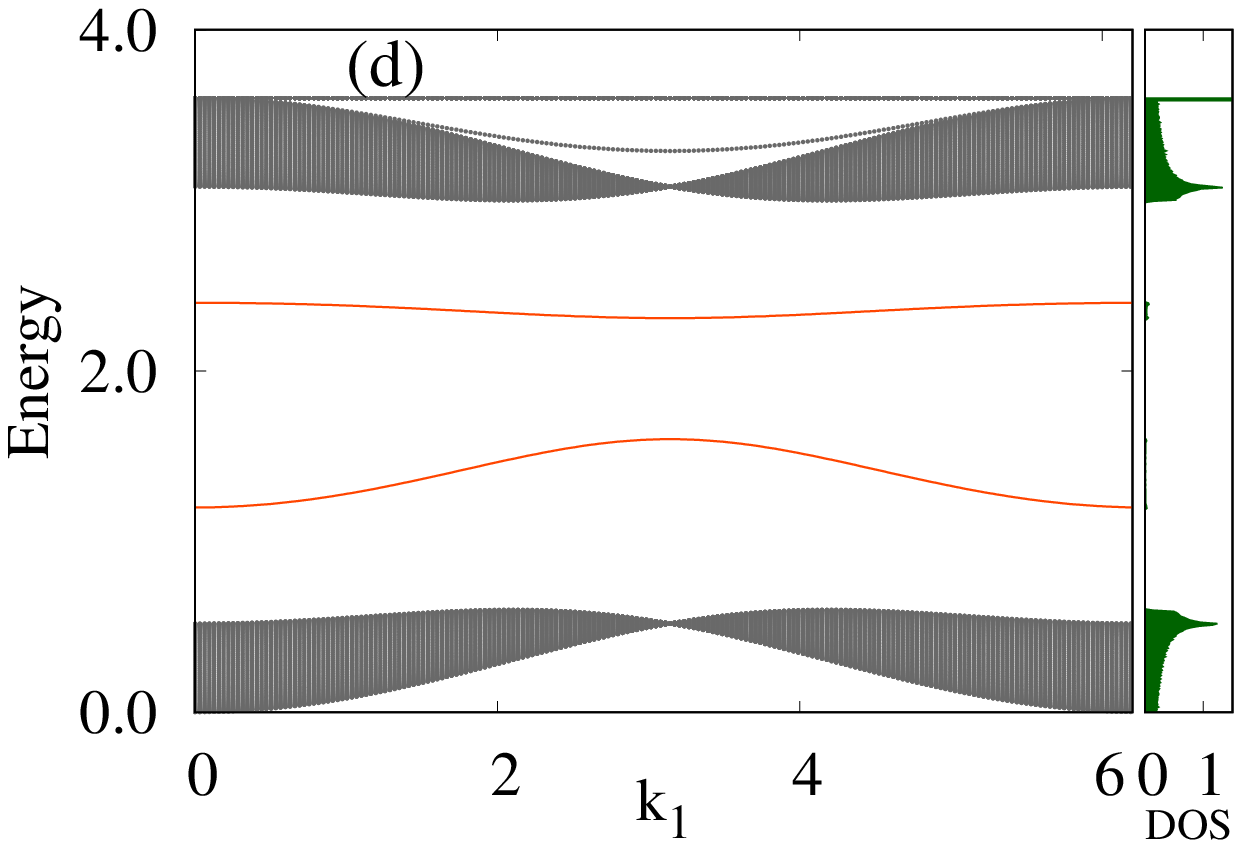}
\includegraphics[width=5cm,height=4cm,trim={0.0cm 0.0cm 0.0cm 0.5cm}]{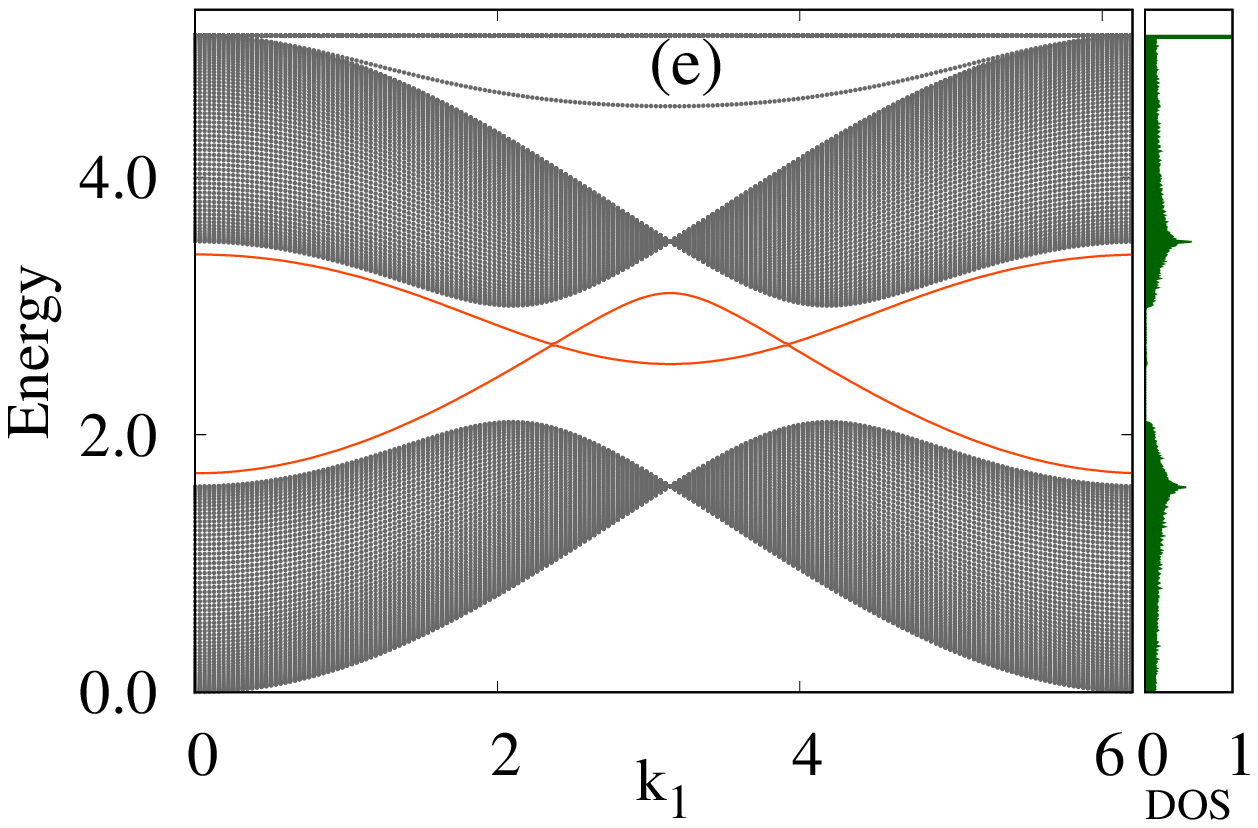}
\includegraphics[width=5cm,height=4cm,trim={0.0cm 0.0cm 0.0cm 0.5cm}]{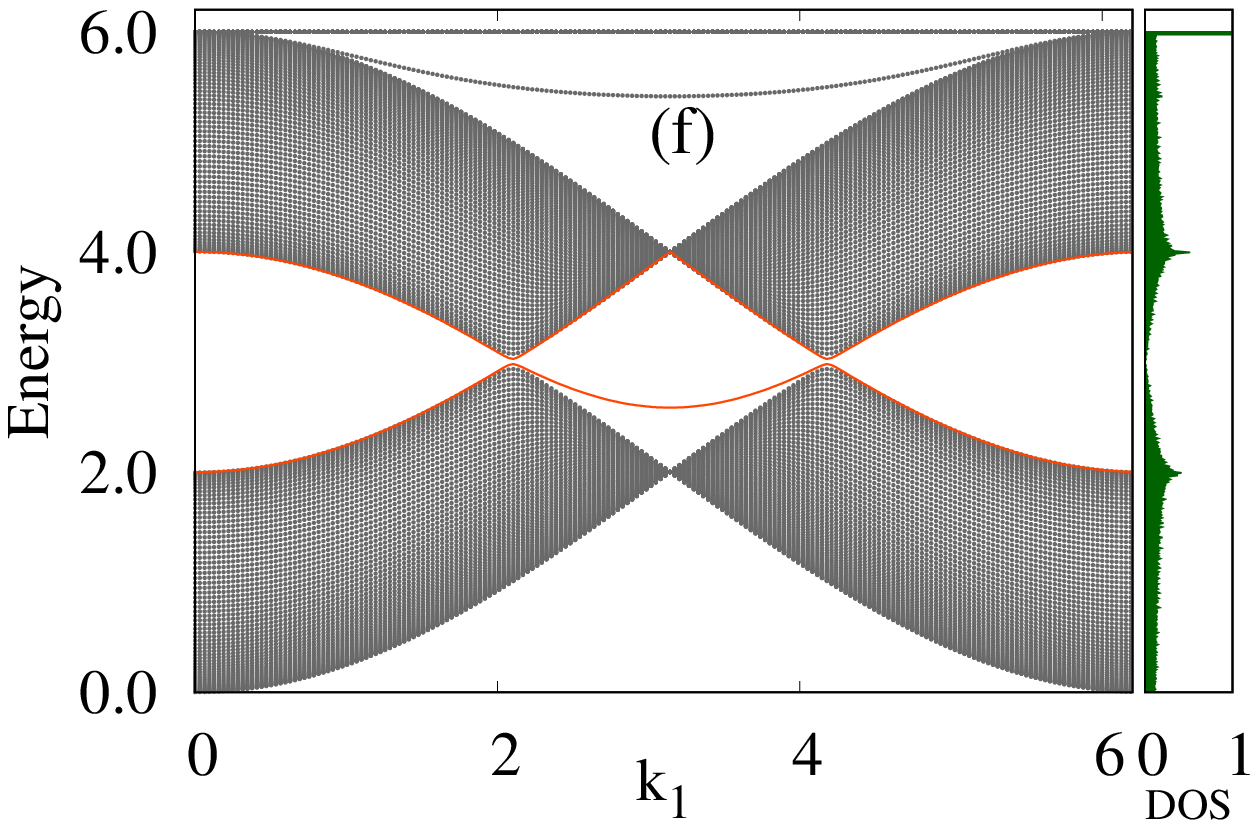}
   \caption{(color online) Dispersion relation of breathing kagome system for zero DMI strength
   with $J_2=1$ and (a)$J_1=0.2$, (b)$J_1=0.7$, (c)$J_1=1.0$.
   Edge state diagram for finite lattice with 70 unit cells along $k_2$ direction
   with same parameter values as the corresponding bulk band structures in the upper row.
   The diagrams (d), (e) and (f) show the evolution of the in-gap edge modes 
   (indicated by red lines)
   and depicts the way it affects the existence of corner states of the system.
   Density of states are shown in the side panel of each edge state diagram.}
 \label{DM0}
  \end{figure*}  
  \section{Topological properties with zero DMI}
  \label{properties}
While kagome ferromagnet ($J_1=J_2$) is topologically trivial in the absence of 
DMI, FM breathing kagome ($J_1 \neq J_2$) with zero DMI is found nontrivial. 
  In this section, topological nature of the system 
will be discussed by studying both bulk and boundary properties of it in terms of suitable 
 topological invariants to characterize them. Henceforth, the value of 
  $J_2$ is fixed at unity while exploring the variation of topological phases 
with respect to the parameter $J_1$. Three sets of bulk dispersion relation 
are shown in Fig \ref{DM0} (a), (b) and (c), where $J_1=0.2$ for the region 
$0.0<J_1<0.5$, 
$J_1=0.7$ for the region $0.5<J_1<1.0$, and $J_1=1.0$, respectively. 
Note that the uppermost band is always 
  flat and it touches the lower band at four corners of the Brillouin zone spanned by
  $(k_1,k_2)$, which are essentially the equivalent points. 
At this moment, the system is an insulator at 1/3 filling
  as the lower two bands are separated. The gap decreases with the increase of $J_1$ and vanishes
  at $J_1=1.0$. The gap again opens up for $J_1>1.0$.
  
  Below the bulk spectrum, we present the band structure of the corresponding finite 
  strip of the system for every case. They are shown in Fig \ref{DM0} (d), (e) and (f). 
The finite strip is prepared by breaking the periodic boundary 
condition (PBC) along the $k_2$ direction.
  In the region, $0.0<J_1<0.5$, two gapped edge modes are found to exist 
   between the lower two gapped bands, which do not decay into the bulk anymore. 
These are the signature of corner states,
  as proved in the previous studies \cite{Ezawa,Sen}. For $0.5<J_1<1.0$, two edge modes
 are found to cross each other twice, without decaying into the bulk again. For $J_1\geq 1.0$, there are no 
  such edge modes. From this edge state spectrum, it is confirmed 
  that the system is topologically trivial in first order since 
there is no gap between the upper two bands
in the region $0.0<J_1<1.0$. Instead, the system is found to host 
a nontrivial second order topological phase in the above 
region.
  \begin{figure*}
  \centering
\includegraphics[width=5cm,height=4.5cm,trim={0.3cm 0.4cm 0.3cm 0.0cm}]{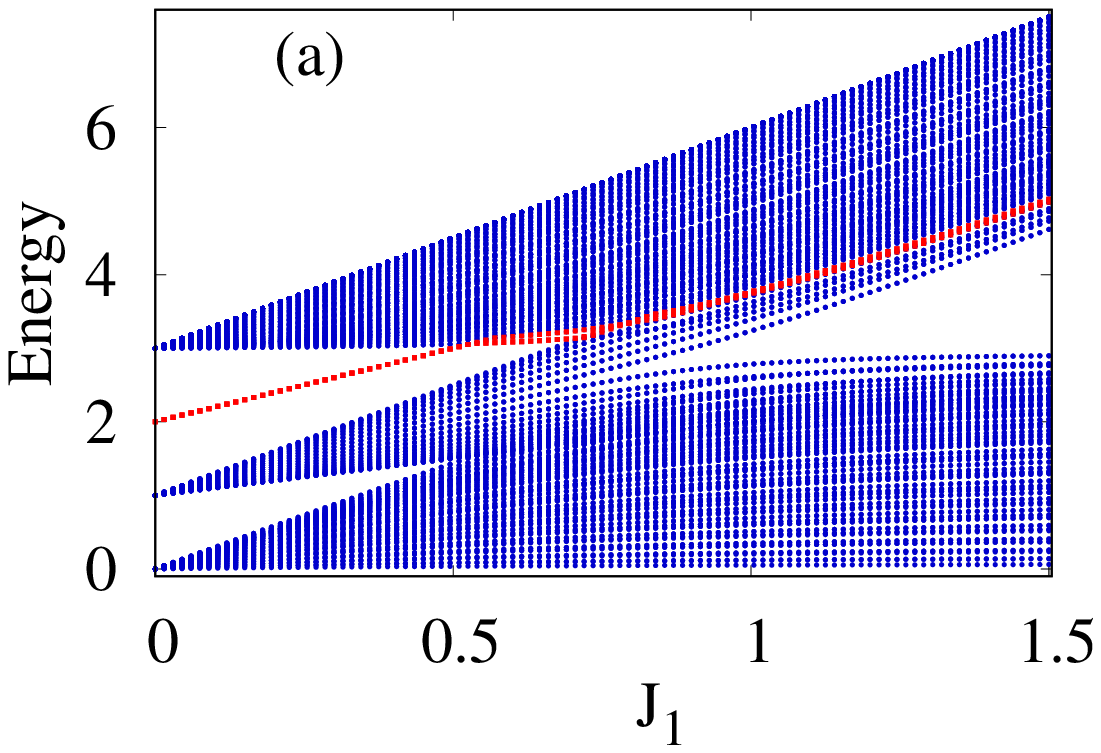}
\includegraphics[width=5cm,height=4.5cm,trim={0.3cm 0.0cm 0.3cm 0.0cm}]{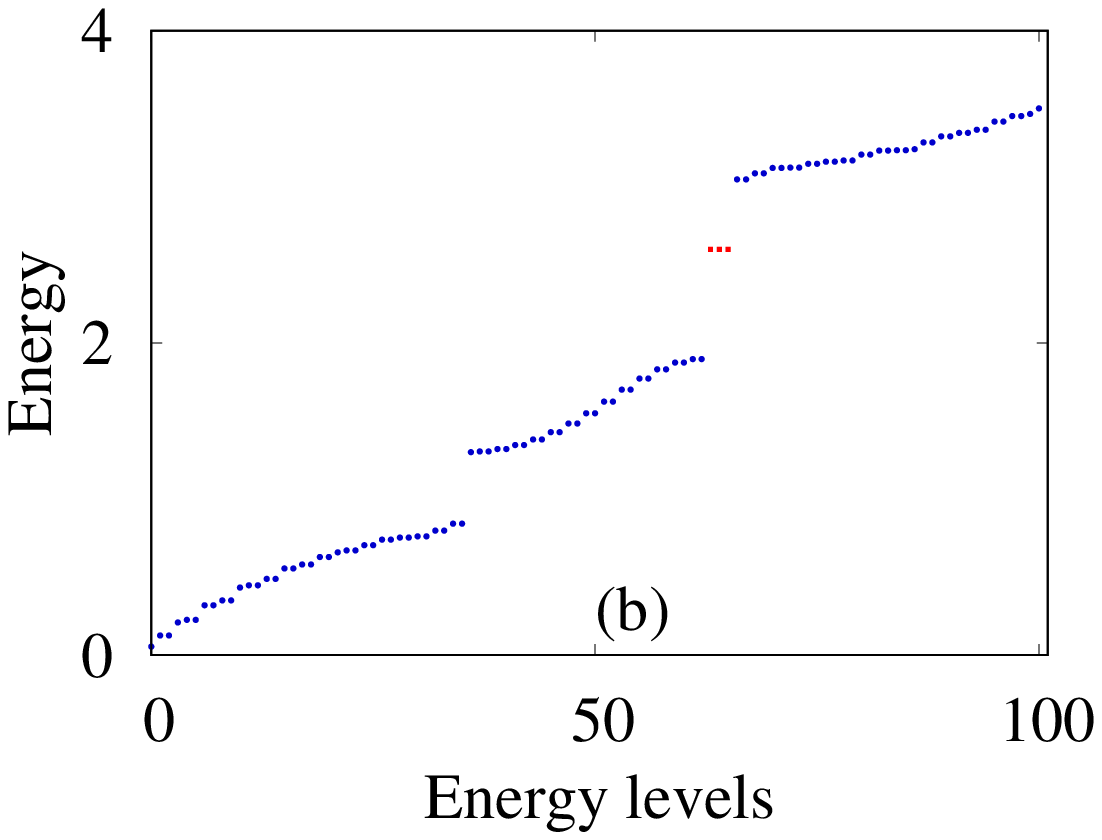}
\includegraphics[width=5cm,height=4.5cm,trim={2.0cm 1.0cm 2.0cm 1.0cm}]{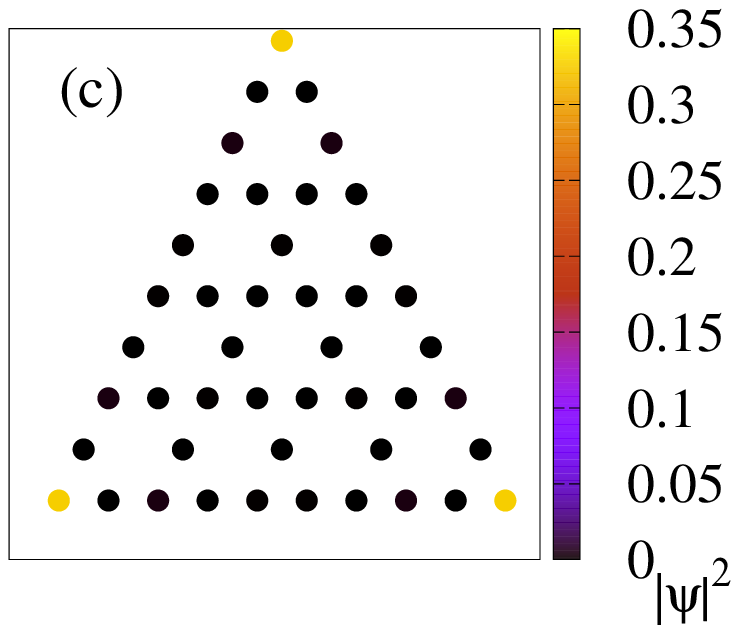}
   \caption{(color online) (a) Energy spectrum of breathing kagome lattice with varying $J_1$
   for $L=15$.
   Upto $J_1=0.5$, zero dimensional corner states (indicated by red dots)
   show its existence and beyond this value,
   it seems to decay into the bulk. (b) Energy of the same system is plotted with 
   respect to energy levels for $J_1=0.3$. Three red dots at same energy, $E=2.6$
   show the states at the three
   corners of the finite triangle. The diagram is truncated to 100 energy levels (instead
   of 164 for $L=10$) for better resolution of the three corner states. (c) Probability
   distribution of a particular eigenstate corresponding to a corner state energy
   for $J_1=0.2$ and $L=5$. It shows that the states are indeed localized at the corners.}
 \label{corner}
  \end{figure*} 
  
  For the characterization of topological phases, bulk topological invariant 
has been formulated by following the procedure developed in the 
  article \cite{Ezawa}. In this
formulation, a particular quantity, polarization along the $n_i$ axis is defined by
  \begin{equation}
 \begin{aligned}
  p_i &=\frac{1}{S}\iint_{\rm 1BZ}d^2\textbf{k} A_i,
    \label{Pi}
 \end{aligned}
\end{equation}
  where $A_i=-i\langle\psi|\partial_{k_i}|\psi\rangle $ is the Berry connection
  with $i=1,2$ and $S=4\pi^2$ being the area of the first Brillouin zone (1BZ)
  spanned by $k_1$ and $k_2$. The set of polarization $(p_1,\,p_2)$ is identical
  to the coordinates of the Wannier
  center \cite{Ezawa}. The distance of the Wannier center from the origin can be taken as the 
  bulk topological invariant as it changes its value only if the gap closes.
  For simplicity, we take $p_1$ as the topological index as it is protected
  by the mirror symmetry along the $n_1$ axis and it is also quantized. As we will see,
  it is non-zero in the topological phase and zero in trivial phase. It can be calculated 
  analytically in extreme cases when either $J_1=0, J_2\neq0,\, (p_1=1/3)$ or 
  $J_2=0, J_1\neq0,\, (p_1=0)$ as for the
  characterization of topological and trivial phases, respectively. 
For example, when $J_1=0, J_2\neq0$
  the exact ground state wave function turns out to be 
$\psi={(1,e^{ik_1},-e^{ik_2})}^T/\sqrt{3}$.
  So, the Berry connection, $A_1$, as well as polarization, $p_1$, 
becomes equal to the value 1/3, following the formula \ref{Pi}. 

Values of $p_i$ have been obtained numerically for every 
non-zero value of $J_1,\,J_2$ and DMI strength. 
To evaluate the integral, Eq \ref{Pi}, we discretize the 
  Brillouin zone and redefine $p_i$ as $p_{in}$ ($n$ being the band index)
  and $A_i$ as $A_{in}=-i\langle\psi_{n\mathbf{k}}|\partial_{k_i}|\psi_{n\mathbf{k}}\rangle $.
 The value of $p_{1n}$ has been calculated for every band, $n$. 
In this case, as the gapped edge states
  exist between the lower two bands, $p_1$ should have a quantized value for 
  the lowest band. Numerical evaluation obtains the value of 
$p_{1}=1/3$ for $0.0<J_1<1.0$, and $p_{1}=0$ for $J_1>1.0$, 
for the lowest band. Hence, the non-zero value of topological
  invariant confirms the nontrivial second order 
topological phase of the system
  for the region mentioned above.
  
  In order to investigate the existence of corner states in this system, we 
  consider a triangular replica of the breathing kagome lattice, as shown in Fig. 
  \ref{lattice}, whose size is defined by the number of small triangles, $L$, along 
  every edge. The triangular replica preserves the three-fold rotation symmetry, $C_3$, 
of the breathing kagome lattice as well as it has the minimum number of 
corners, which is three in this case. 
Different shapes of the finite lattice can be considered for this 
purpose. In Fig \ref{corner} (a), we plot the energy spectrum as a function
  of $J_1$ for $L=15$, which shows that corner states do exist 
for the region $0.0<J_1<0.5$.

  The numerical evaluation of the topological index $p_1$ shows that
  it bears the value 1/3 for the entire region $0.0<J_1<1.0$, which indeed should be the
  case as the insulating phase exists up to $J_1=1.0$ and the invariant has no scope 
  to change its value since no phase transition occurs in the intermediate point. 
On the other hand, corner states cease to show its existence as soon as 
  the gapped edge modes are found to cross each other. And it occurs  
in this system, when $J_1$ 
becomes greater than 0.5, which is illustrated in Fig \ref{DM0} (e). 
As a result, no corner states are found beyond $J_1=0.5$. 
But, the system hosts the second order topological phase
  in the entire region $0.0<J_1<1.0$, even though the 
corner states are topologically 
protected only up to $J_1=0.5$. It implies that bulk-corner-correspondence is  
satisfied for the region $0.0<J_1<0.5$, and violated for 
the region $0.5\leq J_1<1.0$. 
However, the system becomes trivial beyond the limit, $J_1=1.0$.
  In Fig \ref{corner} (b), the energy spectrum is plotted with respect to
  the energy levels for a particular value of $J_1$, which clearly shows the
  existence of three degenerate corner states at energy 2.6$J_2$ 
which correspond to the three different corners of 
  the triangular replica. The number of corner states may vary with the 
shape of replica with different number of corners. 
The distribution 
of probability density for a particular parameter value in this topological phase
   is shown in Fig \ref{corner} (c),  
  which clearly exhibits that the corner states are truly localized at each corner.
    \begin{figure*}
   \centering
\includegraphics[width=5cm,height=6cm,trim={3.0cm 1.0cm 3.0cm 0.0cm}]{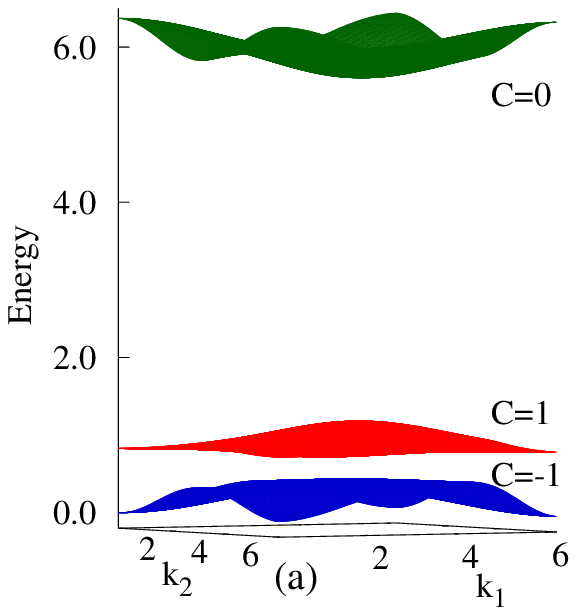}
\includegraphics[width=5cm,height=6cm,trim={3.0cm 1.0cm 3.0cm 0.0cm}]{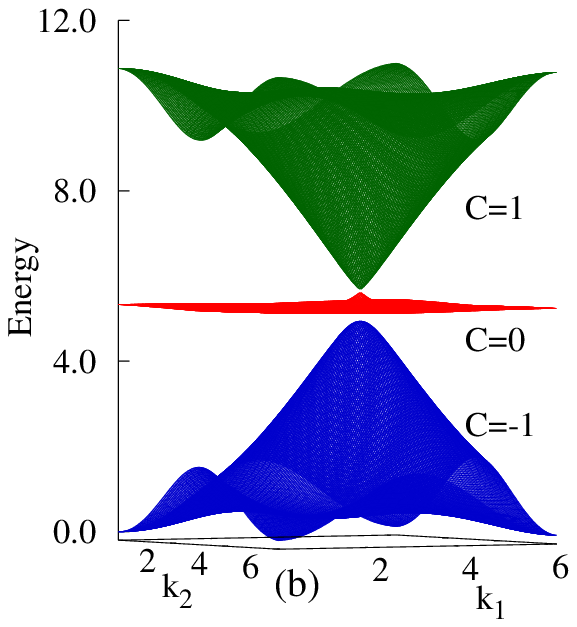}
\includegraphics[width=5cm,height=6cm,trim={3.0cm 1.0cm 3.0cm 0.5cm}]{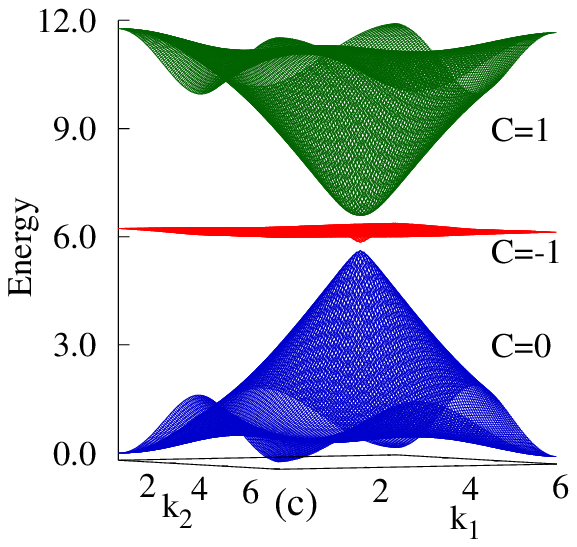}
\includegraphics[width=5cm,height=4cm,trim={0.0cm 0.0cm 0.0cm 0.5cm}]{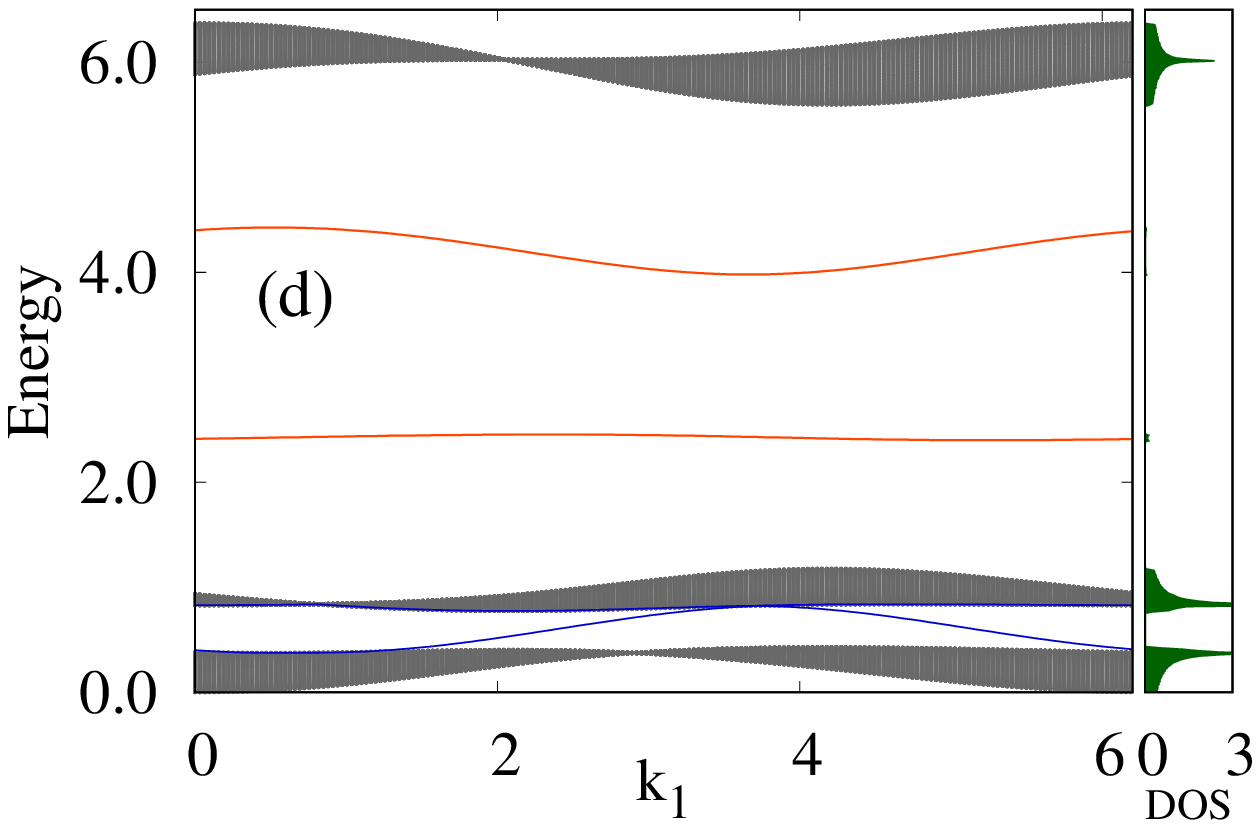}
\includegraphics[width=5cm,height=4cm,trim={0.0cm 0.0cm 0.0cm 0.5cm}]{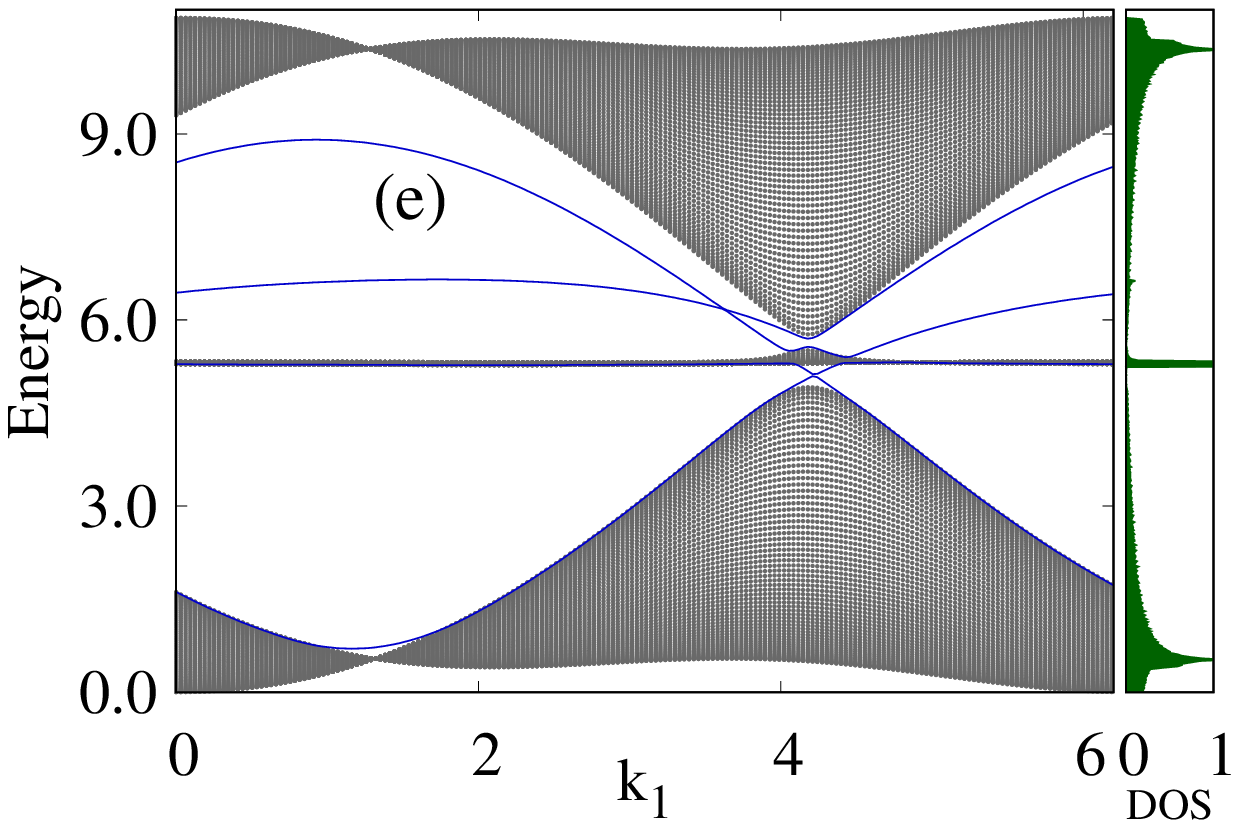}
\includegraphics[width=5cm,height=4cm,trim={0.0cm 0.0cm 0.0cm 0.5cm}]{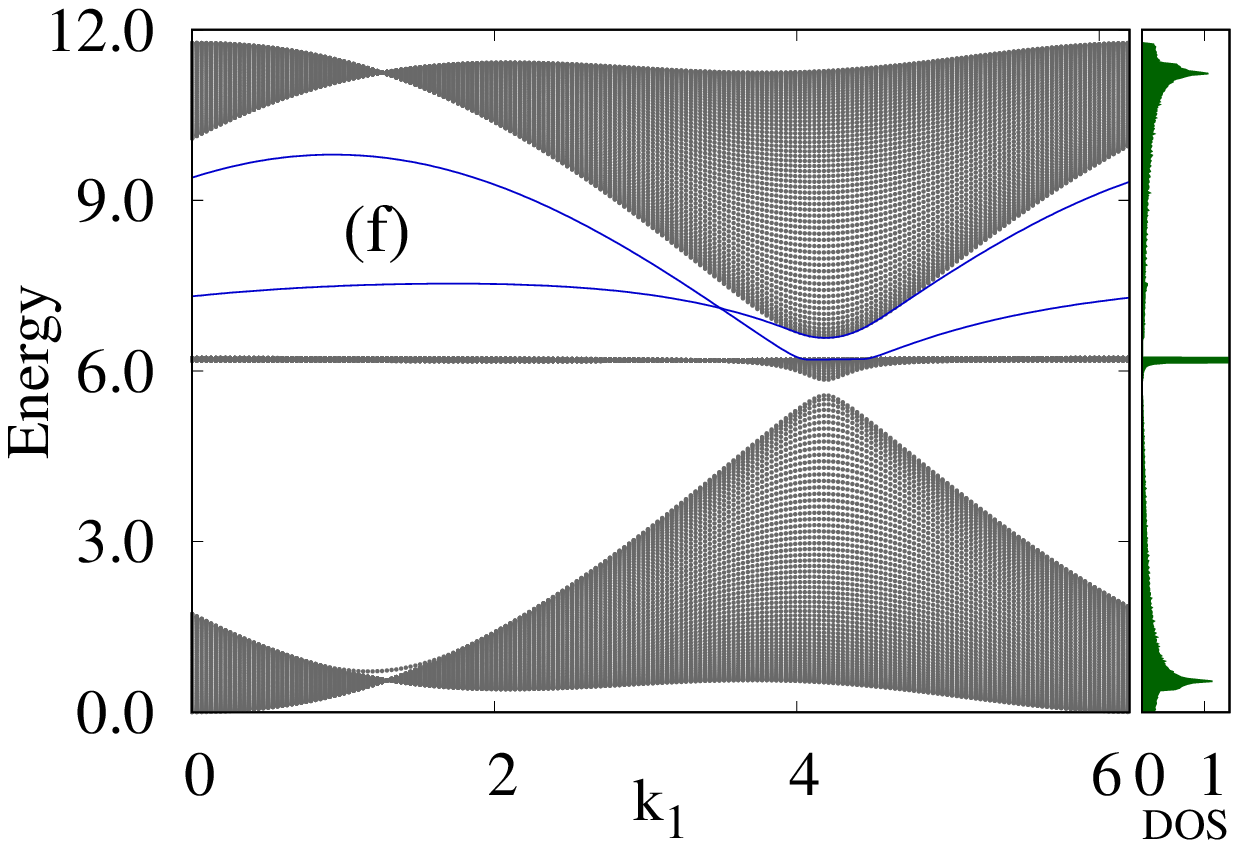}
   \caption{(color online) Dispersion relation of breathing kagome system for non-zero DMI strength
   with $J_2=1.0$, $D_1=0.1$, $D_2=1.5$ and (a) $J_1=0.2$, (b) $J_1=1.7$, (c) $J_1=2.0$. Chern
   numbers of respective bands are also specified.
   Edge state diagram for finite lattice with 70 unit cells along $k_2$ direction
   with same parameter values as the corresponding bulk band structures in the upper row.
   The diagrams (d), (e) and (f) show the evolution of the gapless and gapped edge modes
   and depicts the way it affects the existence of corner states of the system. 
Pair of in-gap 
   edge modes (denoted by red lines) indicates the existence of second order 
   topological phase, while
   chiral edge modes (blue lines) signify the existence of first order topological phase.
   Density of states are shown in the side panel of each edge state diagram.}
 \label{DMI}
  \end{figure*}  
  
  \begin{figure*}
  \centering
\includegraphics[width=5cm,height=4.5cm,trim={0.3cm 0.4cm 0.3cm 0.0cm}]{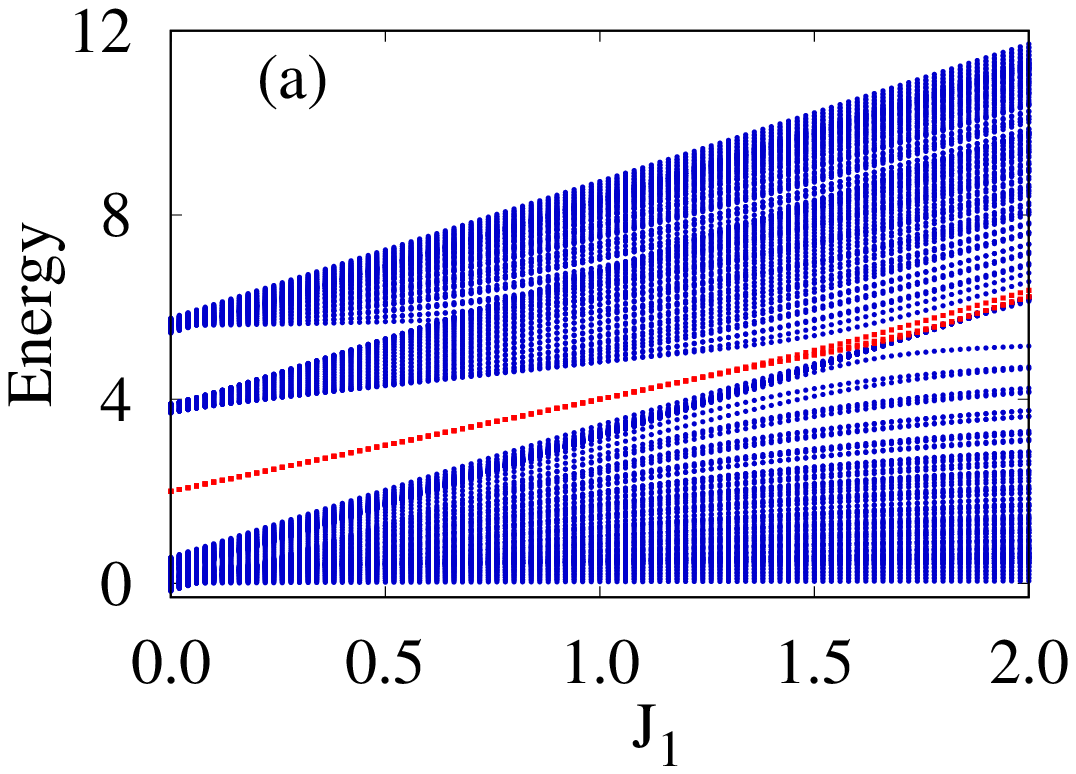}
\includegraphics[width=5cm,height=4.5cm,trim={0.3cm 0.0cm 0.3cm 0.0cm}]{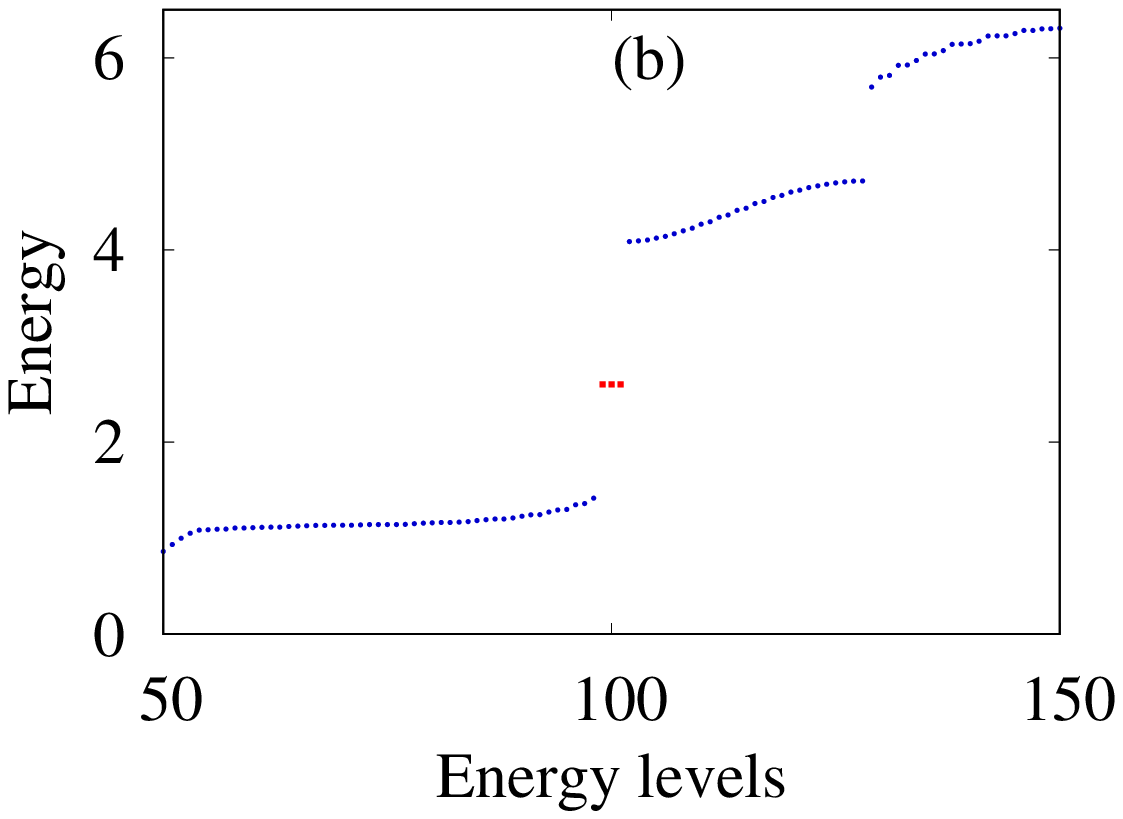}
\includegraphics[width=5cm,height=4.5cm,trim={2.0cm 1.0cm 2.0cm 1.0cm}]{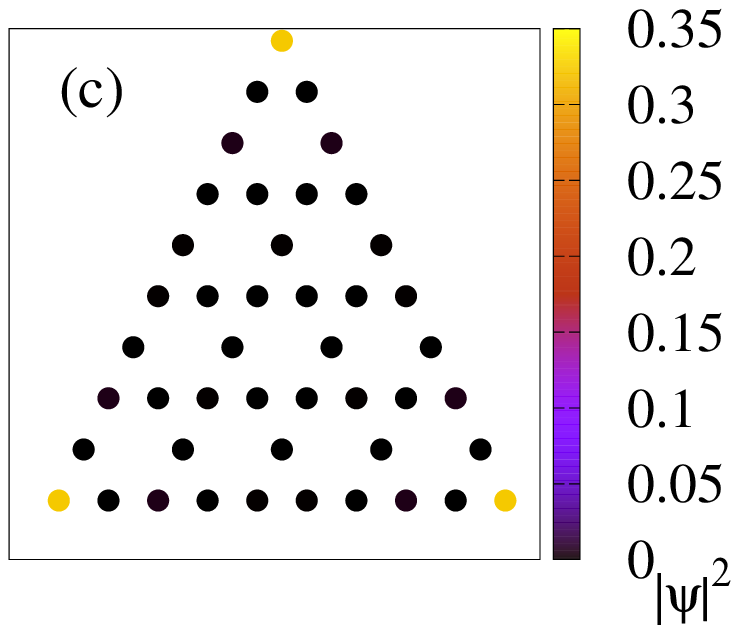}
   \caption{(color online) (a) Energy spectrum of breathing kagome lattice with varying $J_1$
   for $L=15$. $D_1,\,D_2$ and $J_2$ are kept constant at the values of 0.1, 1.5 and 1.0, respectively.
   Upto $J_1=1.6$, zero dimensional corner states exist (denoted by red dots) and beyond this,
   it decays into the bulk. (b) Energy of the same system is plotted with 
   respect to energy levels for $J_1=0.3$. Three red dots at same energy (same value
   as in the case with zero DMI)
   show the states at the three
   corners of the finite triangle. The diagram is truncated 
   for better resolution of the three corner states. (c) Probability
   distribution of a particular eigenstate corresponding to a corner state energy
   for $J_1=0.2$ and $L=5$. It shows that the states are indeed localized at the corners
   and introduction of DM interaction has not changed the distribution considerably.}
 \label{cornerDM}
  \end{figure*} 

  \section{Topological properties with non-zero DMI}
  \label{DM}
The kagome ferromagnet ($J_1=J_2$ and $D_1=D_2$) exhibits an 
unique first order TMI phase in the presence of DMI \cite{Li}. 
On the other hand, FM breathing kagome with non-zero DMI ($J_1 \neq J_2$ and 
 $D_1\neq D_2$) exhibits a rich topological phase diagram which 
includes distinct first and second order TMI phases as well as 
coexistence of both phases. 
In this section, effect of DMI on the topological properties will be discussed. 
DMI is turned on within upward and downward triangles 
with strenghts $D_1$
  and $D_2$, respectively. 
For some specific values of $J_1$, $D_1$ and $D_2$,  
coexistence of both first and second order topological phases is found. 
Two such cases will be described extensively, where
both edge and corner states are found simultaneously. 
Otherwise, distinct first or second order topological phases 
appear in different regions for the nontrivial cases. 

  Now, to characterize the first order topological phase, Chern number 
   $C_n$ for the $n$-th band has been evaluated 
which is defined as the integration of the Berry curvature, 
$\Omega_n (\textbf{k})$, over the 1BZ, {\em i.e.}, 
\begin{equation}
 \begin{aligned}
  C_n &=\frac{1}{2\pi}\iint_{\rm 1BZ}d^2\textbf{k} \Omega_n\left(\textbf{k} \right),
    \label{Cn}
 \end{aligned}
\end{equation}
where $\Omega_n\left(\textbf{k} \right)=
-i\left(\braket{\partial_{1} \psi_{n,\textbf{k}}|\partial_{2} \psi_{n,\textbf{k}}}-
    \braket{\partial_{2} \psi_{n,\textbf{k}}|\partial_{1} \psi_{n,\textbf{k}}}\right)$. 
Here $\ket{\psi_{n,\textbf{k}}}$ are the eigenvectors of $h(\textbf{k})$ 
and $\partial_{i}=\frac{\partial}{\partial k_i}$. In this article, to calculate
the Chern number, we use the 
discretized version of the integration, Eq \ref{Cn}, developed by Fukui and others \cite{Suzuki}. 
   For nontrivial first order topological insulating phase,
  Chern numbers of two or more bands must be non-zero while 
 it will be identically zero for all the bands
for the trivial insulating phase. 
  Chern numbers are undefined when the bands either touch or overlap. 
  To confirm the existence of this first order topological phase, we construct the edge
  state energy diagram by breaking PBC along $k_2$ direction,
  diagonalizing the resulting Hamiltonian and plotting the energy spectrum with 
  respect to the good quantum number $k_1$. 
Likewise, for the characterization of second order topological phase, 
value of polarization is obtained in association with the prediction of 
corner states. To find the corner states
a triangular replica of breathing kagome lattice with $L=15$ 
is considered by breaking PBC along both the directions.  
 
In the presence of DMI, the kagome ferromagnet exhibits a particular TMI phase with 
$C_n=(1,0,-1)$ \cite{Li}. Numbering of band index follows the 
ascending order starting from the lowest energy. 
For FM breathing kagome, 
the uppermost flat band for zero DMI 
is found to become 
dispersive as soon as DMI is non-zero.
The bulk dispersion relations plotted in Fig \ref{DMI} (a), (b) and (c), 
with $D_1=0.1$ and $D_2=1.5$, reveal that the
  system is an insulator for this set of DMI in the 
region $0.00\leq J_1\leq 1.66$, as true gap exists
  between all the bands. At a particular point, $J_1=1.67$, the upper band 
 gap vanishes and reopens thereafter. Therefore 
the system undergoes a phase transition at $J_1=1.67$. 
Similar phase transition occurs at $J_1=1.92$
  when the lower band gap vanishes.
  
  Calculating $C_n$ for each of the cases, it is found that   
$C_n=(-1,1,0)$ for $0.00<J_1\leq 1.66$.
When $J_1\geq1.68$, the Chern numbers are redistributed as $C_n=(-1,0,1)$. 
Thus, the system undergoes topological phase transition through which  
  the upper two bands exchange Chern number of $\pm 1$ since the upper two bands
  touch at a Dirac band touching point. Similar situation happens again
  at $J_1=1.92$, where lower two bands touch at a Dirac band-touching point. 
At this time, they exchange Chern number of $\pm 1$ leading to new distribution, $C_n=(0,-1,1)$.
  
   \begin{figure}[t]
\centering
\includegraphics[width=8cm,height=5.5cm]{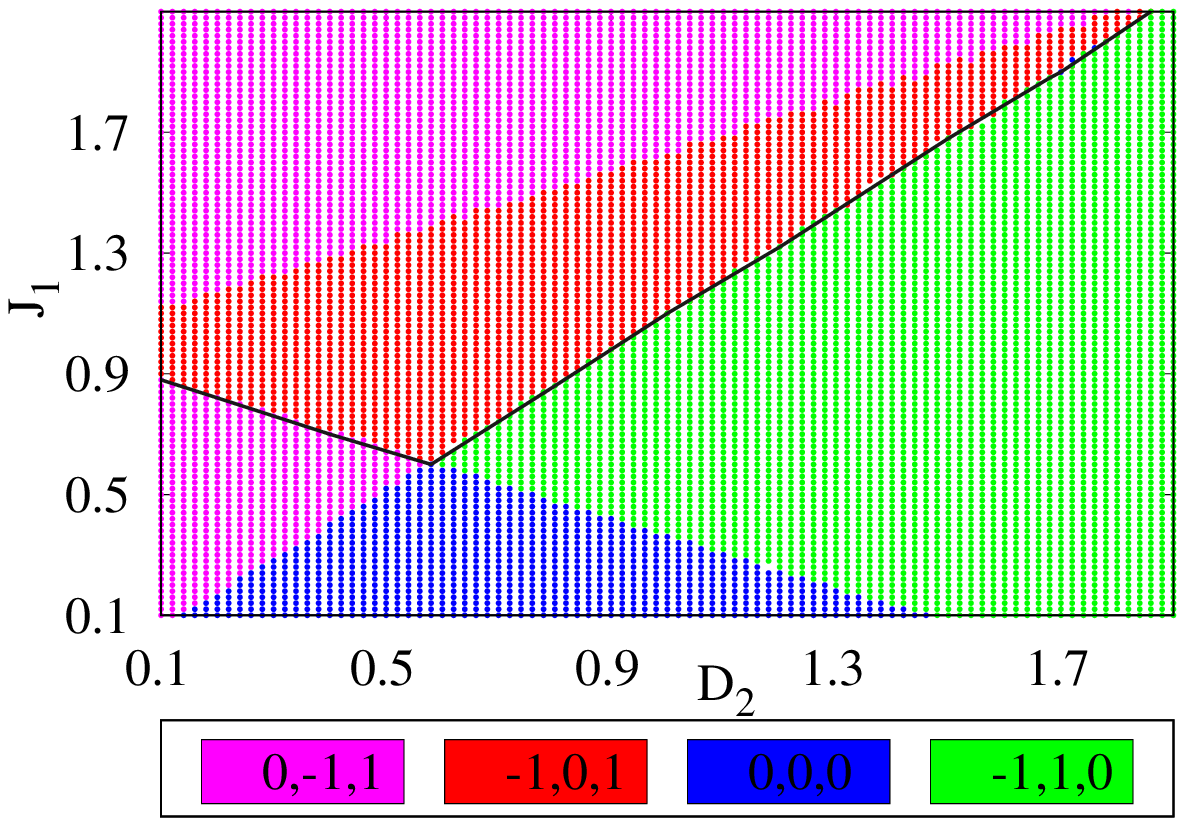}
\caption{(color online) Topological phase diagram in $J_1-D_2$ parameter space
with $J_2=1.0$ and $D_1=0.1$.
Four different phases
are separated by different colors as shown in the lower panel. In first order,
three of them 
are topologically nontrivial and one is trivial. The solid black line
separates topologically nontrivial (lower portion)
and trivial (upper portion) phases in second order.
Evidently, green and magenta portions beneath the solid black line
host both first and second order topological nontrivial phases.
}
\label{TPD}
 \end{figure}
 
  Fig \ref{DMI} (d) shows chiral gapless edge states connecting
  the lower two gapped bands according to the BBC rule 
  since the Chern numbers are $(-1,1,0)$. The pair of gapped edge modes
  in the upper band gap indicates the existence of corner states. 
But this time, they do not cross each other. 
  Thus, there are simultaneous existence of first and second order topological insulating phases 
  for 1/3 and 2/3 filling, respectively.
 In similar fashion, Fig \ref{DMI} (e) and (f), correspond 
  to the existence of other topological phases. 
The nature of edge states supports the pattern 
  of Chern numbers for the corresponding parameter regions satisfying the BBC rule.
  
   To confirm the presence of SOTMI phase, existence of the 
corner states is investigated. 
   Fig \ref{cornerDM} (a) shows the energy spectra with varying $J_1$. It is evident
   that the SOTMI phase do exist up to $J_1=1.66$. 
   The calculation of polarization further emphasizes our claim that TMI
   and SOTMI phase do simultaneously exist in the region $0.00<J_1\leq 1.66$. For this 
   particular case, the topological invariant would be the value of $p_1$ of
   the uppermost band, since the gapped edge mode exist between the upper two bands.
   The value of $p_1$ remains fixed at 1/3 for
   the whole region, which is same as the value of $p_1$ for zero DMI. 
But, in contrast to the zero DMI, here both corner states and non-zero polarization 
simultaneously persists for the region $0.00<J_1\leq 1.66$.
This result can be implied from the fact that there is no crossover 
of in-gap edge modes in this insulating region. 
Thus, BBC as well as bulk-corner-correspondence rules are jointly satisfied 
   both for TMI and SOTMI phases, as evident from the 
   diagram. For $J_1>1.66$, the corner states decay into the bulk 
as well as polarization vanishes. Thus, SOTMI phase cease to exist beyond 
$J_1>1.66$.

This finding clearly predicts the existence of SOTMI phase 
as well as TMI phase in the region $0.00<J_1\leq 1.66$, 
when the strengths of DMI are fixed at $D_1=0.1$ and $D_2=1.5$. 
Therefore, DMI not only helps to extend the range of
SOTMI phase from $0.0<J_1<1.0$ to $0.00<J_1\leq 1.66$, 
in addition, it favors the coexistence of first and 
second order topological phases. 
The system undergoes a phase transition 
in the vicinity of $J_1=1.67$, hosting a new TMI phase 
thereafter.
  
    \begin{figure}[t]
\centering
\includegraphics[width=4cm,height=3.5cm,trim={0.5cm 0.8cm 0.8cm 0.0cm}]{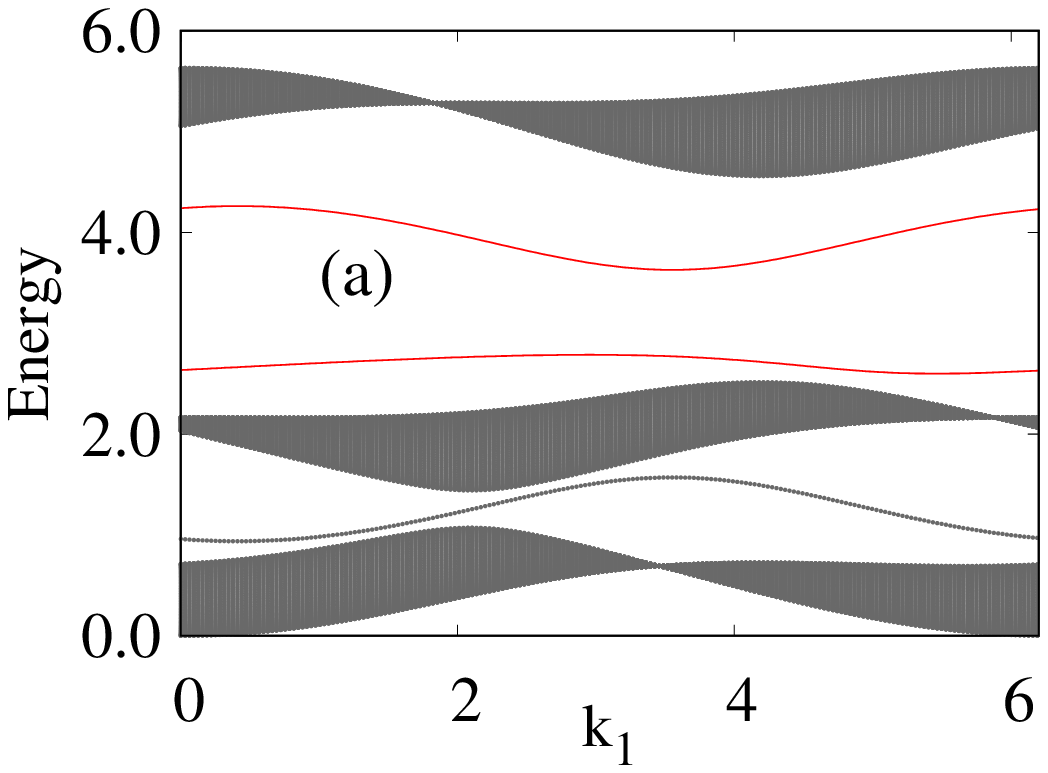}
\includegraphics[width=4cm,height=3.5cm,trim={0.5cm 0.8cm 0.8cm 0.0cm}]{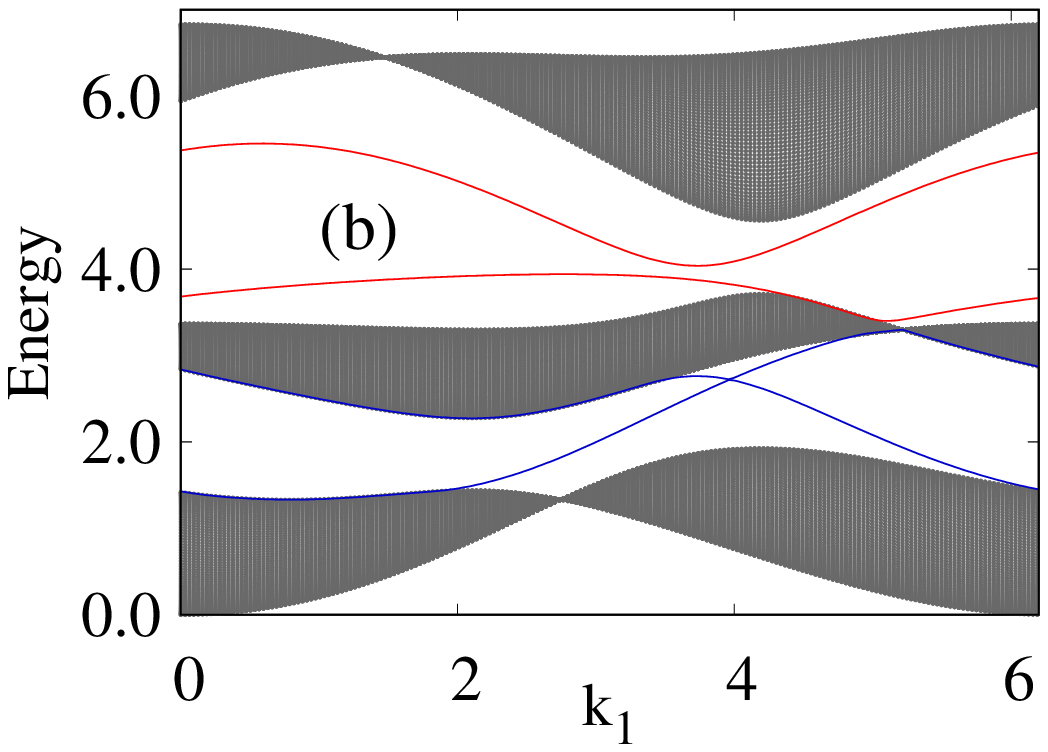}
\caption{(color online) Edge state diagram for $D_1=0.1, D_2=0.9$ and 
(a) $J_1=0.3$, (b) $J_1=0.7$. Upper band-gap hosts two gapped pair of
edge modes (denoted by red lines) in (a), while the lower one decays into
the bulk in (b). A chiral edge state appears between the lower band-gap in (b).
}
\label{edge2}
 \end{figure} 
 A topological phase diagram is presented in Fig \ref{TPD}, which is drawn 
 with respect to two parameters $J_1$ and $D_2$, where $D_1$ is kept fixed at 0.1. 
To explain the phase diagram, let us fix the value of $D_2$ at $0.9$. 
Along this line, the system is topologically
   trivial insulator in first order in the region $0.00<J_1\leq 0.41$. 
Additionally, the system is found to host second order 
topological phase in the same region since 
there is a pair of in-gap edge modes
   in the upper band gap. At $J_1=0.42$ lower gap vanishes. Thereafter, the system hosts a 
TMI phase with $C_n=(-1,1,0)$ up to $J_1=0.97$. Thus, the system becomes topologically nontrivial in first order
for $0.42<J_1\leq 0.97$. But, at the same time, the gapped pair of edge mode in the upper band gap
   changes its shape in such a manner that the lower 
edge mode decays into bulk as it is shown in Fig \ref{edge2}.
   Because of this fact, although the value of 
   $p_1$ for upper band is 1/3 up to $J_1=0.97$, but the corner states are topologically
   protected up to $J_1=0.41$. 
So, violation of bulk-corner-correspondence rule is noted again in the 
region $0.42<J_1\leq 0.97$. 
Thus, for $D_1=0.1$ and $D_2=0.9$,  
 coexistence of both phases remains in the region $0.42<J_1\leq 0.97$. 
For $J_1>0.97$, the upper band-gap vanishes and the system 
undergoes a topological phase transition where it is driven into a phase
   which is topologically trivial in second order but nontrivial in first order. 
   This TMI phase is characterized by the Chern numbers $(-1,0,1)$. 
With further increase of $J_1$, another phase 
transition is observed at $J_1=1.57$. The resulting TMI phase has 
   the Chern number distribution $(0,-1,1)$. 

So, it can be concluded that, although
   polarization, as a topological invariant, changes its value only when
   gap closes, its $(d-2)$ dimensional counterpart, the corner states 
   will be found as long as the pair of in-gap edge modes 
survives distinctly without crossing each other or decaying into 
bulk band in $(d-1)$ dimension. 
   Thus, there is an anomaly in the correspondence 
   between two-dimensional bulk and its zero-dimensional 
boundary for the SOTMI phase.
In addition, this diagram clearly exhibits 
   the occurrence of phase transitions between different topological phases 
with the variation of parameters, $J_1$ and $D_2$. 
  \section{ Thermal Hall Conductivity }
  \label{THCD}
   \begin{figure}[t]
\centering
\includegraphics[width=8cm,height=5cm]{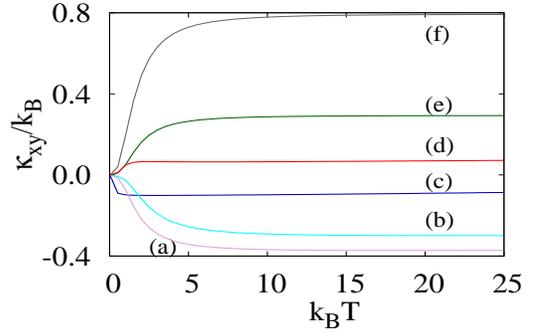}
\caption{(color online) Variation of $\kappa_{xy}(T)$ as a function of T for
$J_2=1.0$ and (a) $J_1=1.7, D_1=0.1, D_2=1.5$, for $C_n=(-1,0,1)$, 
(b) $J_1=2.2, D_1=0.1, D_2=1.5$, for $C_n=(0,-1,1)$,
(c) $J_1=0.2, D_1=0.1, D_2=1.5$, for $C_n=(-1,1,0)$,
(d) $J_1=1.3, D_1=-1.4, D_2=0.3$, for $C_n=(1,-1,0)$,
(e) $J_1=1.3, D_1=0.0, D_2=-0.2$, for $C_n=(0,1,-1)$,
(f) $J_1=1.4, D_1=-0.5, D_2=-0.8$, for $C_n=(1,0,-1)$.
}
\label{THC}
 \end{figure}

   The values of thermal Hall conductivity (THC) of the system
   have been calculated for first order TMI phases. THC is useful to study the 
occurrence of phase transitions, and, at the same time, 
   these values can be verified experimentally.
   Resulting diagram is shown in Fig \ref{THC}. We have included THC values 
for some extra TMI phases those 
are not discussed before. Additional TMI phases with 
different combinations of Chern numbers are obtained 
 by varying all the parameters. The transverse THC
   can be formulated in terms of Berry curvatures, $\Omega_n(\textbf{k})$ as 
\cite{Murakami},
   \begin{equation}
    \begin{aligned}
     \kappa_{xy}(T)=-\frac{{k_B}^2T}{4\pi^2\hbar}\sum_{n}\iint_{\rm 1 BZ} c(\rho_n(\mathbf{k}))dk_xdk_y\Omega_n(\mathbf{k}).
    \end{aligned}
   \end{equation}
   Here the sum runs over all bands, $n$. $k_B$ is the Boltzmann constant and $\hbar$ is the reduced Planck's constant.
   $\rho_n(\mathbf{k})=1/(e^{E_n(\mathbf{k})/k_BT}-1)$ is the Bose distribution function
   with $E_n(\mathbf{k})$ being the energy eigenvalue of the $n$-th band. 
   $c(x)=(1+x)\ln{(\frac{1+x}{x})}^2-(\ln{x})^2-2Li_2(-x)$ where 
   $Li_2(y)=-\int_{0}^{y}dz\frac{\ln{(1-z)}}{z}$. In high temperature limit, THC can be 
   simplified as $\kappa_{xy}=-\frac{k_B}{4\pi^2\hbar}\sum_nC_n\overline{E_n}$ \cite{Cao},
   where the $\textbf{k}$-dependent energy is replaced by the average energy of
   the respective band.
   By using this equation and the distribution of Chern numbers, 
 one can anticipate the sign of saturated value
   of $\kappa_{xy}$ at high temperature. For example,
   sign will be positive (negative) if the band with higher energy has 
   lower (higher) value of Chern number considering one of the bands
    always has $C_n=0$ in this three-band system. 
The behavior of THC is reflected in Fig. \ref{THC}.
 \section{Summary and Discussions}
 \label{summary}
 We have investigated the properties of FM Heisenberg model with and without DMI on 
 breathing kagome lattice and established the simultaneous appearance of first and second order 
 TMI phases for various values of the exchange 
 and DMI strengths which have been taken of different magnitudes for
 upward and downward triangles. 
Topological phase diagram of breathing kagome ferromagnet is 
richer than that of kagome. 
Magnon dispersion relations are obtained
 following LSWT for any value of spin $S$. While only a single SOTMI phase exists when no DMI is 
 present, either SOTMI or TMI or both of them are present for non-zero DMI strength.
 In order to characterize the first order conventional TMI phases the 
 Chern numbers of the insulating band as well as chiral edge states in strip geometry 
are obtained. The existence
 of different phases with different distribution of Chern numbers and the 
 transition between them are studied. Transverse THC values for various 
 TMI phases are also calculated. While SOTMI phases are 
characterized by non-zero values of
 polarization or in terms of the coordinate of Wannier center, those are 
additionally verified by 
 the existence of zero-dimensional corner states where the pair of in-gap edge modes 
are clearly found in one dimension without any crossing. 
 In previous studies on breathing kagome lattice, SOTI phase was found 
 in a fermionic tight binding model \cite{Ezawa}. A TI phase was found on
 another tight binding model in the presence of spin orbit coupling \cite{Nagaosa}.

Six different TMI phases and one SOTMI phase are found 
in this system. 
Since the TMI phase in kagome ferromagnet has been 
observed before in Lu$_2$V$_2$O$_7$ \cite{Tokura,Lee}, 
these findings can also be verified experimentally in future.   
No material is available right now whose property 
can be explained in terms of FM breathing kagome lattice. 
DMI can be induced via external electric field if it is not
 present intrinsically \cite{Cao}. 
Topological phases with higher Chern numbers may be obtained by introducing further neighbor 
interactions. Similarly, it would be more interesting to 
study the topological behavior of the FM models on three-dimensional 
pyrochlore lattice by following the same procedure. 
Violation of bulk-corner-correspondence rule found in some SOTMI phases 
demands more attention as well. Anomaly in bulk-corner-correspondence rule 
is reported before in a Hubbard model on kagome lattice, where gapless 
spin excitations around the corners are found in 
the presence of electron correlations instead of 
gapless charge excitations \cite{Kudo}. 
 
\end{document}